# Micro-modulated luminescence tomography (MLT)

Wenxiang Cong,[1] Fenglin Liu,[1,2] Chao Wang,[1] and Ge Wang[1,*]

[1]Center for Biotechnology and Interdisciplinary Studies, Department of Biomedical Engineering, Rensselaer Polytechnic Institute, Troy, NY 12180, USA

[2]Key Lab of Optoelectronic Technology and System, Ministry of Education, Chongqing University, Chongqing 400044, China

*Corresponding Author: Ge Wang (ge-wang@ieee.org)

**Abstract:** Imaging depth of optical microscopy has been fundamentally limited to millimeter or sub-millimeter due to light scattering. X-ray microscopy can resolve spatial details of few microns deeply inside a sample but the contrast resolution is still inadequate to depict heterogeneous features at cellular or sub-cellular levels. To enhance and enrich biological contrast at large imaging depth, various nanoparticles are introduced and become essential to basic research and molecular medicine. Nanoparticles can be functionalized as imaging probes, similar to fluorescent and bioluminescent proteins. $LiGa_5O_8:Cr^{3+}$ nanoparticles were recently synthesized to facilitate luminescence energy storage with x-ray pre-excitation and the subsequently stimulated luminescence emission by visible/near-infrared (NIR) light. In this paper, we suggest a micro-modulated luminescence tomography (MLT) approach to quantify a nanophosphor distribution in a thick biological sample with high resolution. Our numerical simulation studies demonstrate the feasibility of the proposed approach.

___

## 1. Introduction

Systems biology is devoted to comprehensive studies of biological components with interrelated mechanisms across resolution scales over six orders of magnitude, involving molecules, sub-cellular features, cells, organisms, and entire species [1]. Living systems are highly complicated, dynamic, and often unpredictable. To understand and manipulate these systems, quantitative measurements of interacting components and clusters are necessary using systematic and microscopic technologies such as microscopies, genomics, proteomics, bioinformatics, *in vivo* imaging, and computational models. Regenerative medicine utilizes principles of biology and engineering to develop and transplant engineered substitute tissues and organs [2], with various protocols for cell seeding onto porous scaffolds during incubation [3]. These constructs are then expected to restore or regenerate functionality of diseased tissues or organs. Engineered tissue growths are rather sophisticated, and as natural biological counterparts they usually recapitulate normal developmental processes [4]. Hence, systematic and microscopic technologies are critical for evaluating engineered tissue prior and post implementation.

Molecular and cellular probes have versatile and sophisticated labeling capabilities, and are considered instrumental for systems biology, tissue engineering, and molecular medicine. There is a tremendous interest in biocompatible nanoparticles for *in situ* or *in vivo* molecular imaging, drug delivery, and targeted therapy [5]. Optical imaging is a primary methodology to sensitively visualize nanoparticles tagged to specific molecules and cells [6, 7]. A typical example of their applications is cancer research [8, 9], which employs nanoparticles to deliver drug, heat, or light to cancer cells [10]. Another example is tissue engineering. With multi-functional nanoparticles, engineered tissue constructs can not only be monitored at cellular and molecular levels but also stimulated and regularized by multiple physical means for optimal functionalities. These nanoparticle ingredients are particularly important for the paradigm shift from 2D to 3D matrices in tissue engineering.

Microscopy is the principal observational tool and has made important contributions to our understanding of biological systems and engineered tissues [11]. However, imaging depth of optical microscopy has been fundamentally limited to millimeter or sub-millimeter due to light scattering. Conventional microscopy techniques utilize visible light or electron sources [12-14]. Optical microscopy divides into transmission (*i.e.*, wide-field microscopies for snap-shot of 2D images in light absorption, phase contrast, or dark-field modes) and emission modes (*i.e.*, wide-field fluorescence microscopy, confocal laser scanning microscopy, and two-photon fluorescence microscopy). These microscopic modalities are good for *in vitro* or *in vivo* studies of cultured cell/tissue samples or small animals [15]. Inherently, the resolution of optical microscopy is diffraction-limited by ~200nm with single objective techniques and ~120nm with confocal techniques. With appropriate sample preparation, stochastic information and innovative interference techniques, ~100nm resolution is achievable. Three-dimensional image cubes can be obtained with optical sectioning of ~200nm lateral resolution and ~500nm axial resolution. Ultimately, multiple scattering prevents these techniques from imaging thick samples. Photoacoustic tomography permits scalable resolution at imaging depths up to ~7cm with a depth-to-resolution ratio ~200. Photoacoustic microscopy aims at millimeter imaging depth, micron-scale resolution and absorption contrast, which could be used to characterize the structure of the scaffold but cannot compete with the sensitivity of fluorescence imaging [15, 16].

To enhance biological specifications, various nanoparticles are introduced and become essential to molecular medicine. Nanoparticles can be functionalized as imaging probes, similar to fluorescent and bioluminescent proteins. Unlike conventional nanoparticle probes, $LiGa_5O_8:Cr^{3+}$ nanoparticles were recently synthesized to facilitate the luminescence energy storage with x-ray pre-excitation and the subsequently stimulated luminescence emission by visible/near-infrared (NIR) light.

Our idea is to use a micro-modulated x-ray engraving a distribution of nanophosphors which can then have luminescence energy stored for detailed tomographic imaging deeply in tissue samples. Our goal is to reveal a distribution of $LiGa_5O_8:Cr^{3+}$ nanoparticles targeting specific molecular and cellular aggregates, pathways and responses in engineered tissue samples of several millimeters in size and a few microns in resolution, overcoming the imaging depth limit of all other optical microscopic methods. In this micro-modulated luminescence tomography (MLT) process, tomographic data will be acquired via near-infrared (NIR) light stimulation and optical multiplexing. In the following section, we will describe a system design, a photon transport model, and an image reconstruction algorithm for

MLT. In the third section, we report our realistic numerical simulation results. In the last section, we discuss relevant issues and conclude the paper.

## 2. System Prototyping

**System Architecture**: The proposed MLT system architecture consists of a micro-focus x-ray source, an x-ray zone plate, two EMCCD cameras, NIR laser stimulation sources, and a rotating stage, as shown in Fig. 1. An aluminum filter is used to have a polychromatic x-ray spectrum of 10kev-20kev. As a standard x-ray microscopic imaging component, the zone plate consists of radially symmetric x-ray transparent rings. The width $dr_n$ of a ring decreases with increment of its radius $r_n$. The focal length $f$ of a zone plate is a function of its diameter $D$, its outermost zone width $dr_n$ and the x-ray wavelength $\lambda$: $f = D\,dr_n/\lambda$ [17-19]. When an object intersects an x-ray focal line, the exposed nanophosphors in the object can be excited by x-rays.

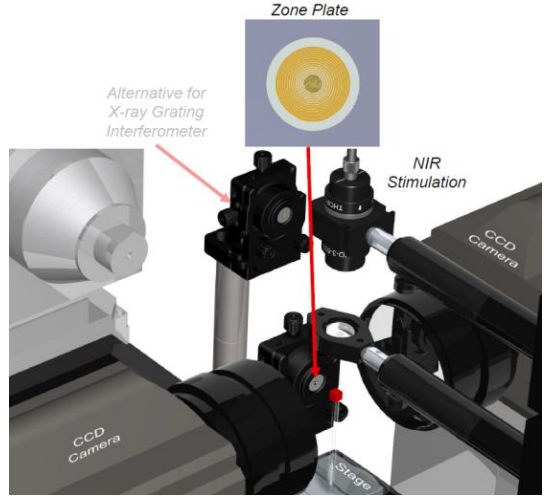

Fig. 1. Modulated luminescence computed tomography system.

Upon NIR light stimulation, the stored x-ray energy will be released by luminescence emission from the x-ray pre-excited nanophosphors to be detected by the CCD camera. All the components will be integrated on an optical table in a light-proof box made of aluminum posts and blackened panels. The x-ray source will be mounted on a horizontally-motorized linear stage (ILS300LM, Newport) for focal plane adjustment. The x-rays can be collimated into a narrow-angle cone-beam to irradiate the Fresnel zone plate. The zone plate is placed within a five axis lens positioner (IP-05A, Newport) to focus polychromatic x-rays to different foci along a line segment for well-localized x-ray excitation of nanophosphors in a phantom or sample. The object to be studied will be placed on a rotational stage (URS75BCC, Newport) on a 3D combination of linear stages (VP-25XA, VP-5ZA, Newport) for x-y-z adjustment. The EMCCD camera (iXon3 897, Andor Technology) has the sensitive 512×512 matrix with pixel size 16×16μm$^2$. The two cameras are faced each other with the object in between for simultaneous data acquisition. NIR laser beams are expanded and controlled to stimulate nanoparticles for luminescence imaging.

## 3. Photon Transport Model and Reconstruction Algorithm

**Photon Transport Model**: MLT involves NIR light stimulation to make energy-stored nanophosphors emit luminescence photons. A light propagation model is needed to describe interactions of light photons with scattering and absorbing media, which is essential for MLT image reconstruction [20, 21]. For biological samples, the diffusion approximation model, a computationally-efficient approximation to the radiative transport equation (RTE), would break down with strong absorbers, near sources, and across boundaries [22, 23]. In that case, either RTE itself or an alternative photon transport model will be needed to accurately describe the photon propagation in biological tissue [23, 24].

For MLT, here we propose an analytic solution of RTE assuming an infinite-space medium. Our solution can be obtained via spherical harmonics approximation [25],

$$\Phi(r) = \sum_{i=1}^{\frac{N+1}{2}} A_i \frac{e^{-v_i r}}{4\pi r} \qquad (1)$$

where $v_i$ and $A_i$ are defined as follows. From the initial conditions $D_0(\lambda)=0$ and $D_1(\lambda)=1$, and the recursive formula $D_{n+1}(\lambda)=(2n+1)\mu_n D_n(\lambda)+\lambda n^2 D_{n-1}(\lambda)$, a polynomial can be derived: $P(\lambda)=D_{N+1}(\lambda)=\sum_{l=0}^{(N-1)/2} a_l \lambda^l$, where $\mu_n = \mu_a + (1-f_n)\mu_s$ and $f_n$ are the n-th order absorption moment and the expansion coefficient of the phase function respectively. In the case of the Henyey-Greenstein phase function, $f_n = g^n$. From the initial conditions $D_0(\lambda)=1$ and $D_1(\lambda)=\mu_a$, the second polynomial can be obtained: $Q(\lambda)=D_{N+1}(\lambda)=\sum_{l=0}^{(N+1)/2} b_l \lambda^l$. The polynomial equation $Q(\lambda)=0$ gives a total of $(N+1)/2$ negative real-valued roots $\lambda_i$. These roots define the values $v_i = \sqrt{-\lambda_i}$ and the coefficients $A_i$

$$A_i = \frac{1}{b_{\frac{N+1}{2}}} \frac{P(\lambda_i)}{\prod_{n=1, n\neq i}^{(N+1)/2}(\lambda_i - \lambda_n)}. \qquad (2)$$

For optical imaging of biological samples, the tissue boundary must be taken into account when analyzing the photon propagation. The luminescence photon propagation in biological tissue is subject to both scattering and absorption. A significant amount of photons go across the tissue boundary and can be detected by a highly sensitive CCD camera. In this scenario, the photon propagation process can be well modeled using a semi-infinite slab. For that purpose, we use the extrapolated boundary condition, which is simple and has been shown to agree well with the MC simulation and physical measurement [26, 27]. An image source is used to construct a fluence rate solution such that $\Phi(x,y,z_z)=0$ holds at an extrapolated boundary at a distance $z_b$ above the surface of the sample, where $z_b = \frac{1+R_{eff}}{1-R_{eff}} \frac{2}{3(\mu_a + (1-g)\mu_s)}$ and $R_{eff} \approx -1.4399n^{-2} + 0.7099n^{-1} + 0.6681 + 0.0636n$ [27, 28]. The photon fluence at the boundary is the sum of the contributions from the source and its image,

$$\Phi(\mathbf{r}) = \sum_{i=1}^{\frac{N+1}{2}} A_i \left( \frac{e^{-v_i r_1}}{4\pi r_1} - \frac{e^{-v_i r_2}}{4\pi r_2} \right), \quad \mathbf{r} \in \partial\Omega, \qquad (3)$$

where $r_1 = \|\mathbf{r}_{src} - \mathbf{r}\|$, and $r_2 = \|\mathbf{r} - \mathbf{r}_{img}\|$. Therefore, the NIR emission from nanophosphors in the sample can be computed using Eq. (3).

**Few-view Image Reconstruction**: The intensity of the NIR luminescence emission depends on the concentration of the nanoparticles $\rho(\mathbf{r})$, the x-ray intensity $X(\mathbf{r})$, the stimulating light intensity $L(\mathbf{r})$, and the luminescence photon yield $\varepsilon$. The luminescence emission from nanophosphors can be formulated as

$$S(\mathbf{r},t) = \eta L(\mathbf{r})\varepsilon X(\mathbf{r})\rho(\mathbf{r})\exp(-L(\mathbf{r})\eta t), \qquad (4)$$

where $\Phi(\mathbf{r})$ is a stimulating light intensity distribution, and $\eta$ is the stimulation efficiency. An x-ray beam can be focused by a zone plate, forming a pair of narrow-angle cones with a common vertex point. When a biological sample is centralized around the focus region, several micrometer width x-rays go through the sample. For example, for a zone plate with a diameter of 0.35mm, an outer zone width of 100nm, and a zone height (gold) of 1600nm, the focal length is 258mm and the maximum width of x-rays through sample is ~3.39μm. In practice, the intensities of x-rays around the focal region can be measured using an x-ray detector, and quantified as the x-ray intensity distribution in the sample. The narrow x-ray beam excites the nanophosphors in the sample, and the luminescence emission from the nanophosphors is measured by the CCD camera. The total intensity of measured NIR light corresponds to the nanophosphor concentration on a narrow x-ray beam path, if we assume that all the nanophosphors in the excited region will be completely depleted during each luminescence data acquisition step (otherwise, the total measurement would be an x-ray intensity weighted integral over the double cone regions). Using the first generation CT scanning mode, we obtained a dataset with a parallel-beam excitation and different view angles for MLT reconstruction. More accurately, we can establish a model to simulate the luminescence emission from nanophosphors. From Eqs. (3)-(4), we can compute the intensity of luminescence emission from the nanophosphors on each x-ray path. Hence, we can establish a system of linear equations:

$$\mathbf{\Phi} = \mathbf{A} \cdot \mathbf{\rho}, \qquad (5)$$

where $\mathbf{\Phi}$ is a vector of measured photon fluence rates, $\mathbf{\rho}$ a vector of nanophosphor concentrations, and $\mathbf{A}$ the system matrix. Then, compressive sensing can be applied to solve the linear inverse problem from few-view data to improve the MLT imaging speed. Specifically, we applied our recently proposed few-view reweighted sparsity hunting (FRESH) scheme [29] for the MLT reconstruction.

**4. Numerical Simulation Results**

Using the MLT methodology, we numerically studied a biological sample phantom of 5×5×5mm³. Biologically relevant optical parameters were assigned to the phantom: absorption coefficient $\mu_a$=0.01mm$^{-1}$, scattering coefficient $\mu_s$=10mm$^{-1}$ and anisotropy parameter g=0.9. A distribution pattern of nanophosphors was embedded in the sample, with 100μg/mL maximum concentration at 2.5mm depth, as shown in Fig. 2 (a). This phantom well describes the nanophosphor accumulation, contrast and spatial resolution. An x-ray source of 5μm focal spot size was filtered by a 0.4mm-thick aluminum plate for an energy spectrum of 10-20KeV. A Fresnel zone plate was utilized for x-ray focusing into a narrow beam. The phantom was steered with 12μm step size and 2.5 ° angular intervals to excite nanophosphors inside the phantom. NIR light stimulation steps were then applied to read out the stored luminescence energy. The x-ray excitation, NIR stimulation and luminescence emission were all simulated according to RTE Eq. (3). The luminescence signals were corrupted by 5% Gaussian noise. In this work, all luminescence photons from a narrow x-ray beam were collected to define the associated line integral, and a sinogram was formed for corresponding reconstruction in a compressive sensing framework. The reconstruction method gave an excellent performance in terms of convergence and stability. The reconstructed images are in a close agreement with the truth, as shown in Fig. 2 (b-c).

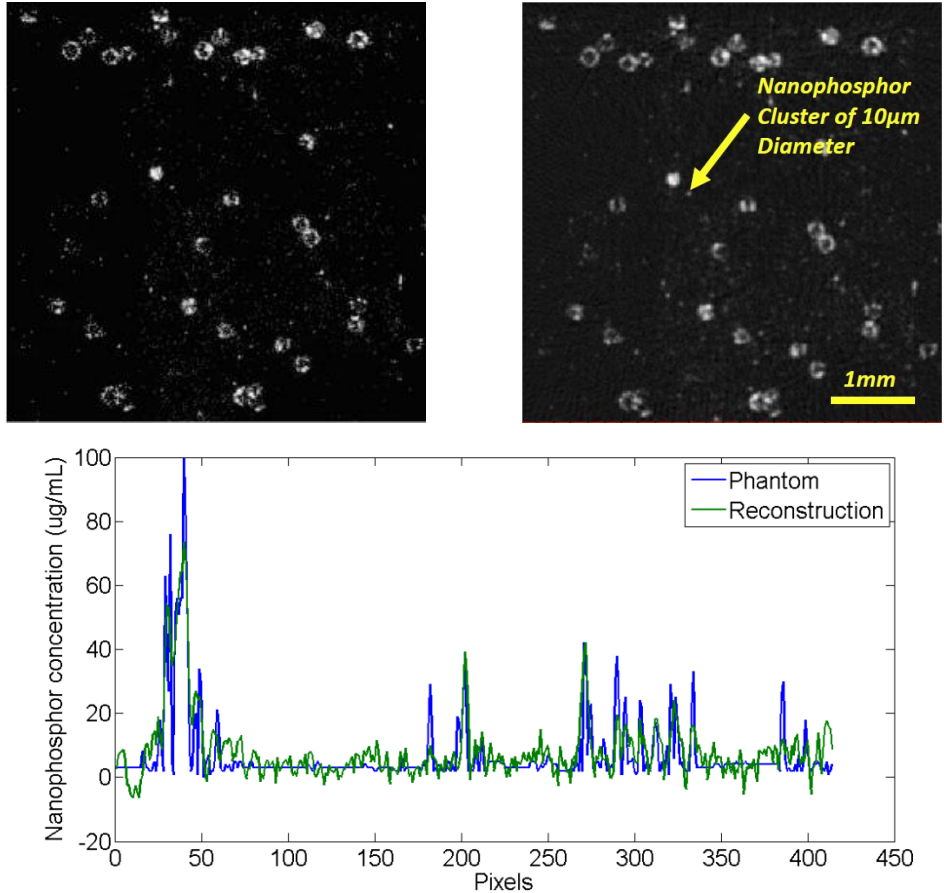

*Fig. 2. Modulated luminescence tomography simulation. (a) The true nanophosphor distribution at 2.5mm depth, (b) the reconstructed nanophosphor distribution at resolution ~10μm, and (c) the profile comparison between true and reconstructed nanophosphors concentration.*

## 5. Discussions and Conclusion

We have proposed the MLT modality to quantify a nanophosphor distribution in a biological sample of a thickness up to 5mm. Different from the conventional pinhole-based x-ray excitation mode adopted for x-ray luminescence tomography [30], which is difficult to manufacture at micro millimeter precision, and such a small pinhole would induce strong diffraction waves, here x-rays are focused to be narrow with a x-ray zone plate or other similar elements such as gratings. A success application of an x-ray zone plate is nano-x-ray computed tomography, which can achieve a nano-scale resolution. Owing to deep penetration of well-modulated x-ray excitation as well as NIR laser stimulation of stored luminescence energy, MLT is capable of achieving micron resolution image reconstruction deep into a sample.

Because of targeted high resolution, this proposed imaging modality will take more imaging time in experiments than x-ray luminescence tomography [30] and stored luminescence tomography [31]. The cutting-edge compressive sensing (CS) reconstruction can recover a sparse image from much fewer measurements than what the Nyquist sampling criterion demands. Few-view tomography techniques will be critical for MLT to reduce projection view and low radiation dose.

In summary, our MLT approach has been formulated. The simulation results have demonstrated the feasibility and potential of the proposed approach. The MLT modality may find many utilities in investigating biological features and processes, monitoring drug delivery, assessing cancer therapy, and other applications.